\documentclass[12pt]{article}%
\usepackage{amssymb}
\usepackage{amsmath}
\usepackage{amsfonts}
\usepackage{graphicx}%
\providecommand{\U}[1]{\protect\rule{.1in}{.1in}}

\begin{document}

\title{Spacial Modulation of the Magnetization in Cobalt Nanowires}
\author{Gerd Bergmann, Yaqi Tao, Jia G. Lu, and Richard S. Thompson\\Department of Physics\\University of Southern California\\Los Angeles, California 90089-0484\\e-mail: bergmann@usc.edu}
\date{\today }
\maketitle

\begin{abstract}
Cobalt nanowires with a diameter in the range between $50$ to $100nm$ can be
prepared as single-crystal \ wires with the easy axis (the c-axis)
perpendicular to the wire axis. The competition between the crystal anisotropy
and demagnetization energy frustrates the magnetization direction. A periodic
modulation of the angle $\theta$ between $\mathbf{M}$ and the wire axis yields
a lower energy.

PACS:

\end{abstract}

In recent years magnetic nanostructures have eperienced a great interest in
dynamic magnetic torque experiments (for ref. see for example \cite{R27}).
These experiments explore the possibility to rotate the orientation of the
magnetization with a current pulse. This could be an important tool in
spintronics. Complementary to the dynamic experiments we want to explore the
static properties of magnetic nanostructures, in particular cobalt nanowires
(NW). We believe that a detailed knowledge of the static magnetic properties
will have important consequences in their dynamic behavior.

A number of experimental groups \cite{B171}, \cite{G52}, \cite{W40},
\cite{G53}, \cite{T18}, \cite{H39}, \cite{G54}, \cite{Z18} have prepared Co
NWs with diameters in the range of $30nm$ to several $100nm$. Similar Co NWs
with a diameter of $80nm$ were recently fabricated at the University of
Southern California \cite{L47}. In some of the experiments the magnetic
structure of the NWs was investigated with a magnetic force microscope (MFM)
\cite{B171}, \cite{G54}, \cite{H39}, \cite{L47}. The MFM scan showed spacial
oscillations of the magnetic field along the length of the wire which are
sometimes quasi-periodic. Thiaville et al. \cite{T19} concluded that in their
experiments the period is in agreement with a "head-to-head" magnetization, a
180$^{\text{o}}$ Bloch wall.

Henry at al. \cite{H39} observed by means of dark field transmission electron
microscopy (TEM) images that the Co NWs have the bulk hexagonal structure. For
wire diameters $2R<50nm$ the easy c-axis lies parallel to the wire axis while
for NWs with diameters of $2R>50nm$ the easy c-axis is perpendicular to the NW
axis. In the following we discuss the latter case, NWs with a diameter
$2R>50nm$. Below we choose a radius of $R=40nm$ for quantitative calculations.
We denote the wire axis as the z-direction and the easy axis as the
x-direction of our coordinate system.

When the axis of the Co NW and the easy axis lie perpendicular to each other
then the magnetization is frustrated. The shape or demagnetization anisotropy
prefers to align the magnetization in the z-direction, parallel to the wire
axis. But the crystal anisotropy definitely favors the x-direction. And this
crystal anisotropy is very large in the uniaxial Co wire.

The shape or demagnetization anisotropy energy density (ED) is due to the
demagnetization field and given by
\[
u_{d}=\frac{\mu_{0}}{2}\mathbf{MNM}=-\frac{\mu_{0}}{2}\mathbf{H}%
\cdot\mathbf{M}%
\]
where $\mathbf{N}$ is the 3x3 demagnetization matrix, $\mathbf{H}$ is the
demagnetization field, i.e. the magnetic field in the absence of an external
magnetic field, and $\mathbf{M}$ the magnetization. We introduce $\theta$ as
the angle between the z- or wire axis and the magnetization $\mathbf{M}$.
(Within this paper the magnetization will always lie in the x-z-plane). Then
one has a demagnetization factor of $N_{xx}=\frac{1}{2}$ for $\theta=\pi/2$
(perpendicular to the wire) and $N_{zz}=0$ for $\theta=0$ (parallel to the
wire axis). For a constant magnetization $\mathbf{M=M}_{0}\left(  \sin
\theta,0,\cos\theta\right)  $ under the angle $\theta$ the demagnetization
energy density is
\[
u_{d}=\frac{1}{2}\sin^{2}\theta\left(  \frac{\mu_{0}}{2}M_{0}^{2}\right)
\]
We take from O'Handley \cite{O33} the reference value for the magnetic ED of
Co $u_{00}=\frac{\mu_{0}}{2}M_{0}^{2}=12\times10^{5}J/m^{3}.$ The value of
$u_{d}/u_{00}$ is $0$ for $\mathbf{M}$ parallel to the wire axis and $1/2$ for
$\mathbf{M}$ parallel to the easy axis.

The energy density of the crystal anisotropy is generally given in terms of
the angle between the easy axis and the magnetization. In our geometry this
angle is equal to $\left(  \frac{\pi}{2}-\theta\right)  $. The crystal
anisotropy ED is, in terms of this angle $\theta$%
\[
u_{ca}=k_{1}\cos^{2}\theta+k_{2}\cos^{4}\theta
\]
The crystal anisotropy constant $k_{1}$ is given in the literature as
$k_{1}=4.1\times10^{5}J/m^{3}$ \cite{O33}. For the constant $k_{2}$ one finds
different values in the literature, for example $k_{2}=1.5\times10^{5}J/m^{3}$
\cite{O33} and $k_{2}=1.0\times10^{5}J/m^{3}$ \cite{C27}. The resulting
crystal anisotropy ED $u_{ca}/u_{00}$ is $\allowbreak0.47$ ($0.425\,)$ along
the wire axis and $0$ parallel to the easy axis. (The value in parenthesis is
for $k_{2}=1.0\times10^{5}J/m^{3}$). This difference in the constant $k_{2}$
has important consequences. The value of $k_{2}=1.5\times10^{5}J/m^{3}$ yields
the lowest ED $\left(  u_{d}+u_{ca}\right)  /u_{00}=$\thinspace$0.450$ for a
finite angle of $\theta=0.65\equiv37^{\text{o}}$ between the magnetization and
the wire axis. For the value of $k_{2}=1.0\times10^{5}J/m^{3}$ the
magnetization would align parallel to the z-axis.

Obviously the competition between the crystal anisotropy and demagnetization
is a close call. The system will try to reduce its energy as much as possible
by the crystal anisotropy without paying too much energy to the
demagnetization energy. One way to reduce the demagnetization energy is to
modulate the magnetization direction in the x-z-plane so that the angle
$\theta$ between $\mathbf{M}$ and $\widehat{\mathbf{z}}$ oscillates as
$\theta=\theta_{0}\cos\left(  kz\right)  $. (There is no oscillation in time
but only in space in contrast to spin waves in NWs which have been treated by
Arias and Mills \cite{M82}). While for a constant magnetization in x-direction
the field $\mathbf{H}$ falls off as $1/\rho^{2}$ with the distance $\rho$ from
the wire axis, a modulated magnetization with a period $\lambda$ will cancel
the field for distances $\rho$ which are larger than $\lambda$. This reduces
the demagnetization ED. In this paper we investigate the effect of such a
modulation on the ED of the wire. This modulation corresponds to a
magnetization $\mathbf{M}$
\begin{equation}
\mathbf{M}=M_{0}\left(  \sin\left(  \theta_{0}\cos kz\right)  ,0,\cos\left(
\theta_{0}\cos kz\right)  \right)  \label{M3d}%
\end{equation}
In Fig.1 the orientation of the magnetization is shown as a function of $z$.
We keep the absolute value of $\left\vert \mathbf{M}\right\vert =M_{0}$ constant.

$%
{\includegraphics[
height=1.3674in,
width=3.428in
]%
{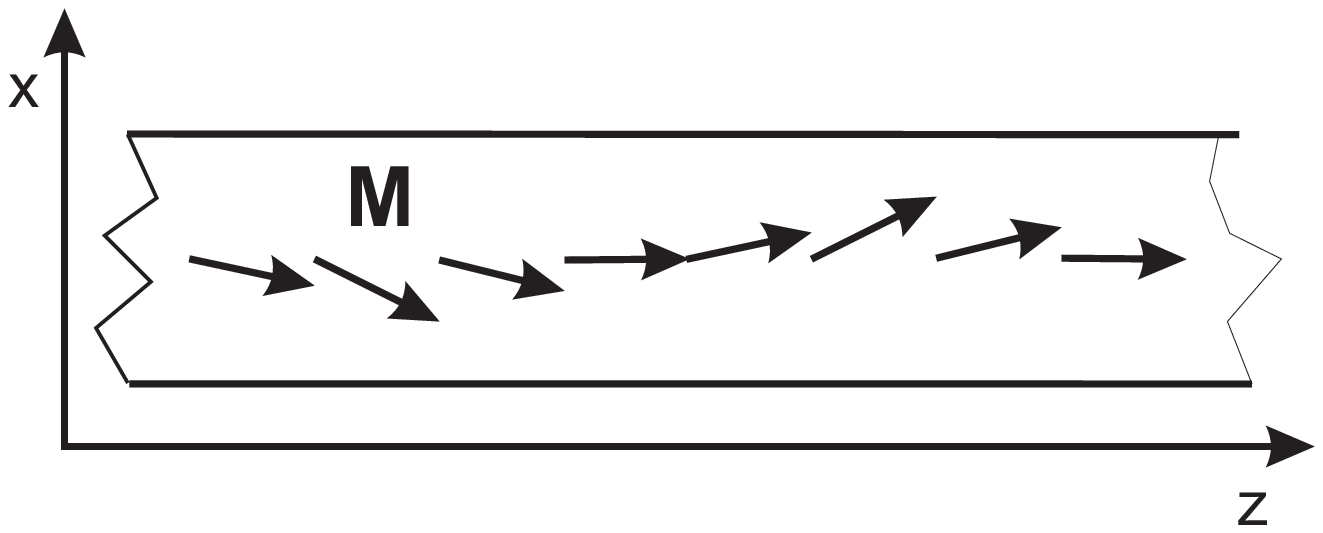}%
}%
$%

\begin{tabular}
[c]{l}%
Fig.1: Nanowire with magnetization modulation along the z-axis.
\end{tabular}%
\[
\]

The magnetization components $M_{x}$ and $M_{z}$ can be expressed as two
Fourier series.
\[
M_{x}\left(  z\right)  =M_{0}%
{\textstyle\sum_{\nu=0}^{\infty}}
c_{2\nu+1}\cos\left[  \left(  2\nu+1\right)  kz\right]
\]%
\[
M_{z}\left(  z\right)  =M_{0}%
{\textstyle\sum_{\nu=1}^{\infty}}
c_{2\nu}\cos\left(  2\nu kz\right)
\]
The coefficients $c_{2\nu+1},c_{2\nu}$ can be easily obtained from a Fourier
expansion of $\mathbf{M}$ in equ. (\ref{M3d}) . The lowest coefficients are
$c_{0}\left(  \theta_{0}\right)  =\left(  1-\frac{1}{4}\theta_{0}^{2}+\frac
{1}{64}\theta_{0}^{4}-+..\right)  ,$ $c_{1}\left(  \theta_{0}\right)  $
$=\left(  \allowbreak\theta_{0}-\frac{1}{8}\theta_{0}^{3}+\frac{1}{192}%
\theta_{0}^{5}-+..\right)  $, etc. We include terms up to the order of
$\theta_{0}^{18}$.

In the next step we calculate the demagnetization field $\mathbf{H}$ for a
magnetization $M_{x}=M_{x0}\cos\left(  qz\right)  $. Setting afterwards
$q=\left(  2\nu+1\right)  k$ and $M_{x0}=M_{0}c_{2\nu+1}$ the results can be
used for each Fourier component.

The magnetic flux $\mathbf{B}$ inside and outside of the sample is given by
$\mathbf{B}=\mu_{0}\left(  \mathbf{H+M}\right)  $. Since there are no external
currents in our problem the curl of the magnetic field vanishes,
$\triangledown\times\mathbf{H}=0$. Therefore the magnetic field can be
expressed as the gradient of a magnetic potential $\mathbf{H}=-\triangledown
\phi$ (in full analogy the electrostatic case). Taking the divergence of the
magnetic flux (which vanishes) yields%
\[
0=\triangledown\cdot\mathbf{B}=\mu_{0}\left(  \triangledown\cdot
\mathbf{H+}\triangledown\cdot\mathbf{M}\right)
\]
and replacing the field by the potential yields
\[
\Delta\phi\mathbf{=}\triangledown\cdot\mathbf{M}%
\]
For $M_{x}$ the divergence of $\mathbf{M}$ is zero.

We use cylindrical coordinates $\left(  \rho,\varphi,z\right)  $ and take the
$\varphi$-dependence as $\cos\varphi$. Then the solutions of the LaPlace
equation are%
\[
\phi=\left\{
\begin{array}
[c]{ccc}%
C^{in}I_{1}\left(  q\rho\right)  \cos\varphi\cos qz &  & \rho<R\\
C^{out}K_{1}\left(  q\rho\right)  \cos\varphi\cos qz &  & \rho>R
\end{array}
\right\}
\]
where $I_{1}\left(  s\right)  $ and $K_{1}\left(  s\right)  $ are modified
Bessel functions. The coefficients $C^{in},C^{out}$ are obtained by using the
boundary conditions at $\rho=R$. The components $B_{\rho}$ and $H_{\varphi}$
have to be continuous. This yields $C^{in}=RM_{x0}K_{1}\left(  qR\right)  $
and $C^{out}=RM_{x0}I_{1}\left(  qR\right)  $. (In determing the coefficients
one obtains the Wronski determinant $W=\left[  I_{1}\left(  qR\right)
K_{1}^{\prime}\left(  qR\right)  -I_{1}^{\prime}\left(  qR\right)
K_{1}\left(  qR\right)  \right]  $ as a denominator, which has the value
$W=-1/\left(  qR\right)  $).

From the magnetic potential one obtains the components of the magnetic field
$\mathbf{H}$. The x-component of $\mathbf{H}$ inside the wire is
\[
H_{x}\left(  \rho<R\right)  =-qRK_{1}\left(  qR\right)  \left[  I_{1}^{\prime
}\left(  q\rho\right)  \cos^{2}\varphi+\frac{1}{q\rho}I_{1}\left(
q\rho\right)  \sin^{2}\varphi\right]  M_{0x}\cos\left(  qz\right)
\]
The local demagnetization ED is $-\left(  \mu_{0}/2\right)  H_{x}M_{x}$. We
average over a period in z-direction and the cross section $\pi R^{2}$ and
obtain for an individual Fourier component the demagnetization ED%

\[
\left(  \frac{\mu_{0}}{2}M_{x0}^{2}\right)  \frac{1}{2}K_{1}\left(  qR\right)
I_{1}\left(  qR\right)
\]

For each $q=\left(  2\nu+1\right)  k$ the demagnetization field $\mathbf{H}$
interacts only with the magnetization $\mathbf{M}$ of the same $q$ (after
averaging). Then the total contribution of all components of $M_{x}$ is just
the sum of the individual contributions. In the following we normalize all EDs
by dividing by the value $u_{00}=\frac{\mu_{0}}{2}M_{0}^{2}$. Then the
normalized ED is
\[
\frac{u_{x}\left(  s,\theta\right)  }{u_{00}}=%
{\textstyle\sum_{\nu=0}^{n}}
\left(  c_{2\nu+1}\left(  \theta\right)  \right)  ^{2}\frac{1}{2}K_{1}\left[
\left(  2\nu+1\right)  s\right]  I_{1}\left[  \left(  2\nu+1\right)  s\right]
\]
where $s=kR$. In the numerical evaluation we include three terms (the third
hardly contributes).

The Fourier components $M_{z}=M_{z0}\cos qz$ for the z-component of the
magnetization are calculated quite analogously. The main difference is that
the magnetic field $\mathbf{H}$ and therefore the magnetic potential are
independent of $\varphi$. Therefore $\phi$ is given by the modified Bessel
functions $I_{0}\left(  q\rho\right)  $ and $K_{0}\left(  q\rho\right)  $ of
order zero. Furthermore $\Delta\phi$ does not vanish but is given by%

\[
\Delta\phi\mathbf{=}\frac{dM_{z}}{dz}=-M_{z0}q\sin\left(  qz\right)  \neq0
\]

The solution is found in complete analogy to the $M_{x}$-component and is
given by%

\[
\phi\left(  \rho,z\right)  =RM_{z0}\sin\left(  qz\right)  \left\{
\begin{array}
[c]{ccc}%
\left[  \frac{1}{qR}+K_{0}^{\prime}\left(  qR\right)  I_{0}\left(
q\rho\right)  \right]  &  & \rho<R\\
I_{0}^{\prime}\left(  qR\right)  K_{0}\left(  q\rho\right)  &  & \rho>R
\end{array}
\right\}
\]
The magnetic field component $H_{z}$ inside the wire is
\[
H_{z}\left(  \rho<R\right)  =-\left(  qRK_{0}^{\prime}\left(  qR\right)
I_{0}\left(  q\rho\right)  +1\right)  M_{0,z}\cos qz
\]
In the evaluation of the demagnetization ED we use the identities
$tI_{0}\left(  t\right)  =d\left(  tI_{1}\left(  t\right)  \right)  /dt,$
$K_{0}^{\prime}\left(  t\right)  =-K_{1}\left(  t\right)  $, $I_{0}^{\prime
}\left(  t\right)  =I_{1}\left(  t\right)  $. The averaged demagnetization ED
becomes $\left(  \frac{\mu_{0}}{2}M_{x0}^{2}\right)  \left(  \frac{1}{2}%
-K_{1}\left(  qR\right)  I_{1}\left(  qR\right)  \right)  $. The contribution
of all Fourier components of $M_{z}$ is
\[
\frac{u_{z}\left(  s,\theta_{0}\right)  }{u_{00}}=%
{\textstyle\sum_{\nu=1}^{\infty}}
\left(  c_{2\nu}\left(  \theta\right)  \right)  ^{2}\left(  \frac{1}{2}%
-K_{1}\left(  2\nu s\right)  I_{1}\left(  2\nu s\right)  \right)
\]
Again we include the first three terms in the numerical evaluation.

Next we consider the crystal anisotropy ED. The average of the term $k_{1}%
\cos^{2}\theta$ yields%
\[
\frac{u_{ca}^{\left(  1\right)  }\left(  \theta_{0}\right)  }{u_{00}}=\frac
{1}{u_{00}}\frac{1}{2\pi}\int_{0}^{2\pi}k_{1}\cos^{2}\left(  \theta_{0}%
\cos\left(  s\right)  \right)  d\left(  s\right)  =0.34\times a_{1}\left(
\theta_{0}\right)
\]
where $a_{1}\left(  \theta_{0}\right)  =1-\frac{1}{2}\theta_{0}^{2}+\frac
{1}{8}\theta_{0}^{4}-+..$. The average of the term $k_{2}\cos^{4}\theta$
yields%
\[
\frac{u_{ca}^{\left(  2\right)  }}{u_{00}}=8.\,\allowbreak3\times10^{-2}\times
a_{2}\left(  \theta_{0}\right)
\]
for $k_{2}=1.5\times10^{5}J/m^{3}$ with $a_{2}\left(  \theta_{0}\right)
=1-\theta_{0}^{2}+\frac{5}{8}\theta_{0}^{4}-+..$. \ In both cases we include
terms up to the order of $\theta_{0}^{18}$.

Finally we have to include the exchange stiffness of the Co wire. While a
modulation of the magnetization can reduce the demagnetization and the crystal
anisotropy energy, it will cost energy because the of the bending of the
magnetization. The increase in the ED can be expressed in terms of the
exchange stiffness constant $D_{ex}$%

\[
u_{ex}=\frac{1}{4}\frac{M_{0}}{g\mu_{B}}\theta_{0}^{2}D_{ex}k^{2}%
\]
Since the energy densities $u_{x}$ and $u_{z}$ are a function of $kR$ and
$\theta_{0}$ we express all energies as functions of $s=kR$ and $\theta_{0}$.
Then we obtain
\[
\frac{u_{ex}}{u_{00}}=\frac{1}{4}\frac{M_{0}D_{ex}}{u_{00}R^{2}g\mu_{B}}%
\theta_{0}^{2}s^{2}%
\]
Liu et al. \cite{S76} determined the exchange stiffness $D_{ex}$
experimentally from the spin-wave spectrum in hexagonal Co. They also
performed a theoretical calculation. From the experiment they obtained
$D_{ex}=435meV\times A^{2}=6.96\times10^{-40}Jm^{2}$. Their theoretical result
yielded twice this value. Using the experimental value and a radius of
$R=40nm$ we obtain $u_{ex}$ $=8125$ $\ast\left(  kR\right)  ^{2}\theta_{0}%
^{2}$ $\left[  Jm^{-3}\right]  $. The normalized exchange stiffness ED is
then
\[
\frac{u_{ex}}{u_{00}}=a_{ex}s^{2}\theta_{0}^{2}\text{, }a_{ex}=0.6\,8\times
10^{-2}%
\]
This exchange ED is very small compared with the demagnetization and the
crystal anisotropy EDs which are of the order of 1.

Finally we add all terms and calculate the total ED as a function of $s=kR$
and $\theta_{0}$ and determine the minimum of this energy
\[
u_{t}\left(  s,\theta\right)  =\frac{1}{u_{00}}\left[  u_{x}\left(
s,\theta_{0}\right)  +u_{z}\left(  s,\theta_{0}\right)  +u_{ca}\left(
\theta_{0}\right)  +u_{ex}\left(  s,\theta_{0}\right)  \right]
\]

We perform the calculation for different choices of the parameter $k_{2}$ and
determine the position of the minimum of the ED in the $s$-$\theta_{0}$-plane.
To investigate the effect of the exchange ED we also perform a calculation
with twice the experimental value for $a_{ex}$. In table I the numerical
results for different parameter choices are collected.%

\begin{align*}
&
\begin{tabular}
[c]{|l|l|l|l|l|l|l|}\hline
$\mathbf{k}_{2}/u_{00}$ & $\mathbf{a}_{ex}/u_{00}\left[  10^{-2}\right]  $ &
$\mathbf{s}_{\min}$ & $\theta_{\min}$ & $\mathbf{u}_{\min}/u_{00}$ &
$\mathbf{u}\left[  \mathbf{M||}\widehat{\mathbf{z}}\right]  $ & $\mathbf{u}%
\left[  \mathbf{M||}\widehat{\mathbf{x}}\right]  $\\\hline
$0$ & $0.6\,8$ & $2.3$ & $0.7$ & $0.333\,88$ & $\allowbreak0.342$ &
$.5$\\\hline
$0$ & $1.\,\allowbreak36$ & $1.75$ & $0.3$ & $0.341\,37$ & $\allowbreak0.342$
& $.5$\\\hline
$.083$ & $0.6\,8$ & $2.1$ & $1.0$ & $0.378\,83$ & $0.425\,$ & $.5$\\\hline
$.083$ & $1.\,\allowbreak36$ & $1.6$ & $0.8$ & $0.396\,9$ & $0.425\,$ &
$.5$\\\hline
$0.125$ & $0.68$ & $2.1$ & $1.0$ & $0.39704$ & $\allowbreak0.425\,$ &
$.5$\\\hline
$0.125\,$ & $1.\,\allowbreak36$ & $1.5$ & $0.9$ & $0.417\,71$ & $\allowbreak
0.425\,$ & $.5$\\\hline
\end{tabular}
\\
&
\begin{tabular}
[c]{l}%
Table I: For two values of $k_{2}$ and $a_{ex}$ the coordinates and the value
of the\\
(normalized) energy density (ED) in the $s-\theta_{0}$-plane are collected in
columns\\
three, four and five. Columns six and seven give the ED for a constant\\
magnization parallel to the z- and the x-asis.\\
\end{tabular}
\end{align*}

For $k_{2}/u_{00}=0.125$ and $a_{ex}=0.68\times10^{-2}$ \ we find the minimum
at $\left(  s,\theta_{0}\right)  $ $=\left(  2.1,1.0\right)  $. In Fig.2a,b
the dependence of $u_{t}/u_{00}$ is plotted for these parameters. The figures
show two orthogonal traces through the energy minimum (a) along\ the $s=kR$
direction and (b) along the $\theta_{0}$ direction.

$%
{\includegraphics[
height=2.1934in,
width=2.7248in
]%
{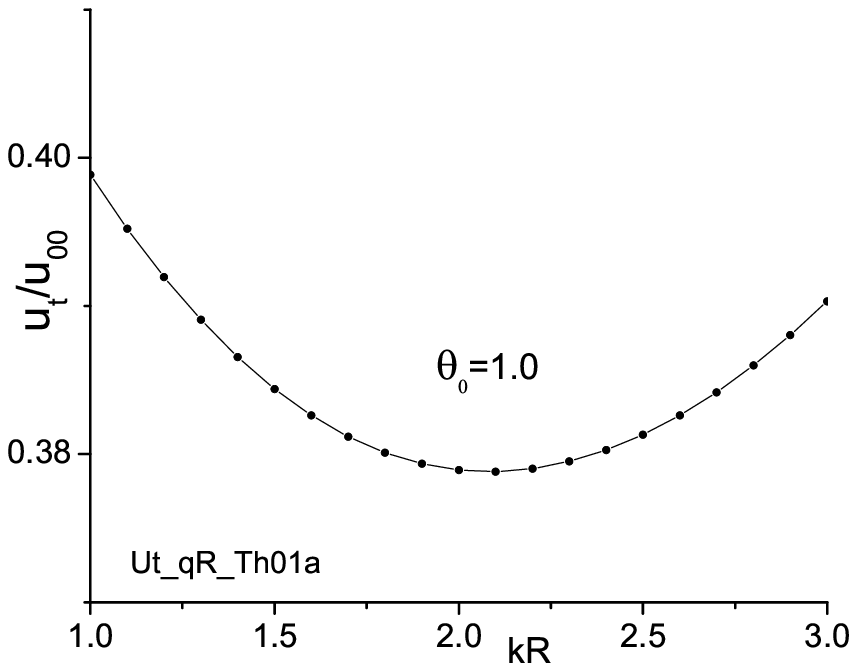}%
}
{\includegraphics[
height=2.1851in,
width=2.6891in
]%
{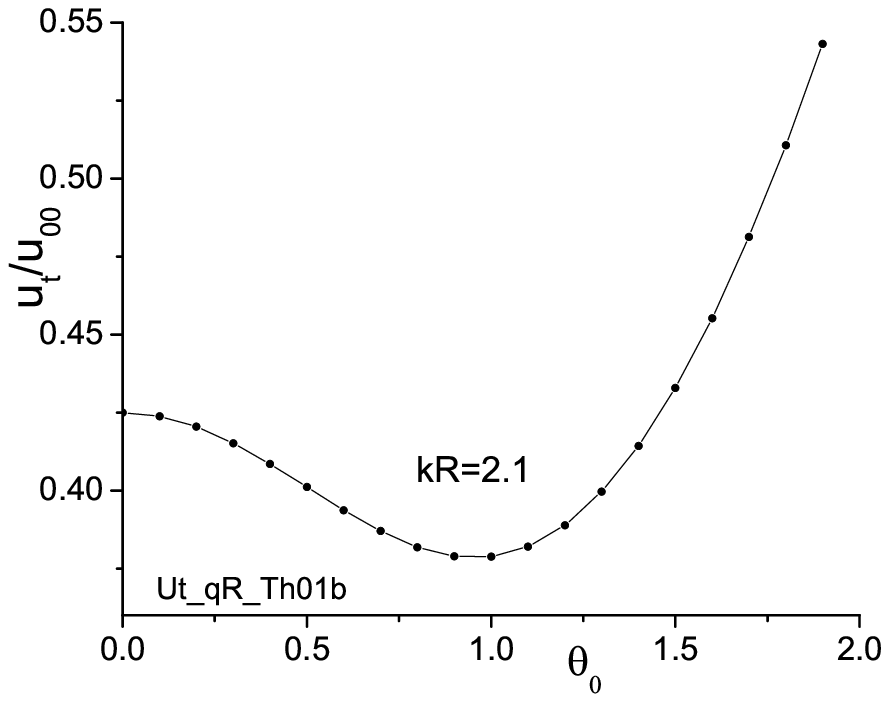}%
}
$

$%
\begin{tabular}
[c]{l}%
Fig.2a,b: The ED as a function of $s=kR$ (3a) and $\theta_{0}$ (3b) through\\
the minimum for the parameters $k_{2}/u_{00}=0.125$ and $a_{ex}/u_{00}%
=0.6\,8\times10^{-2}$.
\end{tabular}
\ \ $%
\[
\]

For $\theta_{0}=1.0$ we can draw the two components $M_{x}$ and $M_{z}$ as a
function of $z$ along the wire. This is shown in Fig.3. With $\theta
=1.0\ast\cos\left(  kz\right)  $ the amplitude of the angle is less than
$\pi/2.$ Therefore the \ z-component never reverses direction. At $\sin\left(
1.1\right)  =0.84$ the x-component reaches almost the saturation magnetization.

$%
{\includegraphics[
height=2.1959in,
width=2.7215in
]%
{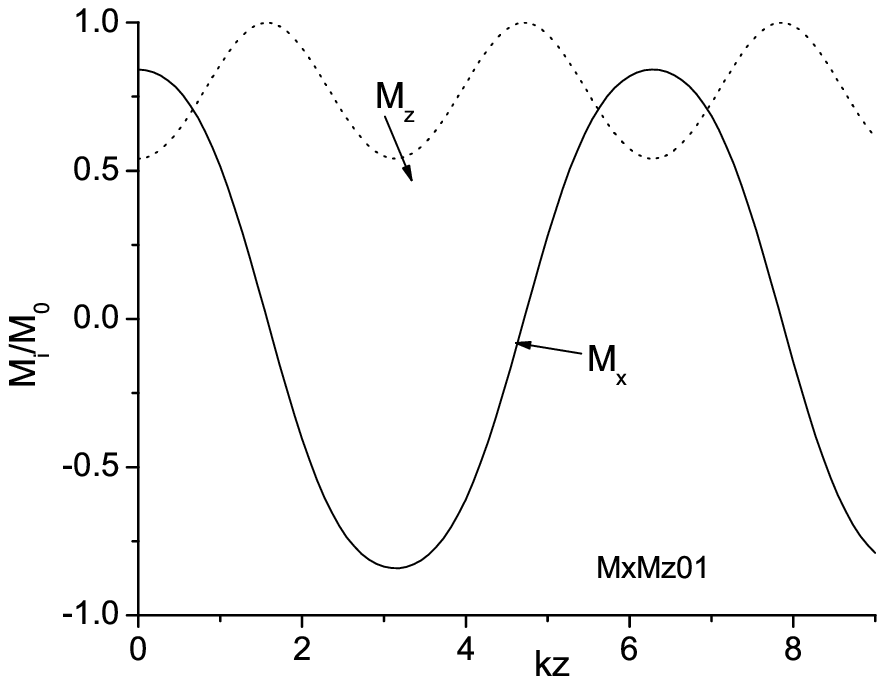}%
}
$

$%
\begin{tabular}
[c]{l}%
Fig.3a: The x- and the z-component of the\\
magnetization as a function of position $s=kz$.
\end{tabular}
$%
\[
\]
$\ $

For a comparison we calculate the ED when the magnetization angle rotates in
the x-z-plane as $\theta=\cos\left(  kz\right)  $. This yields
\[
\mathbf{M=}M_{0}\left(  \sin\left(  \kappa z\right)  ,0,\cos\left(  kz\right)
\right)
\]
In this case we have only one Fourier component in x- and z-directions with
the same wave number $k$. The demagnetization ED follows from the above
calculation. (There is no cross term between the x- and z-part of the
demagnetization ED since their $\varphi$-components are orthogonal). The
$k_{1}$-part of the crystal anisotropy ED has the weight $1/2$ and the $k_{2}$
has a weight of $3/8$. The exchange stiffness ED is just $u_{ex}%
/u_{00}=0.68\times10^{-2}s^{2}$. Fig.4 shows the dependence of the total ED
$u_{t}/u_{00}$ as a function of $s=kR$. The total ED has its smallest value of
$u_{t}/u_{00}=0.467\,7$ at $k=0$. This value is considerably higher than for a
constant magnetization along the NW axis with $u_{||}/u_{00}=0.408$. Therefore
this behavior of the magnetization is energetically unfavorable.

$%
{\includegraphics[
height=2.3669in,
width=2.9049in
]%
{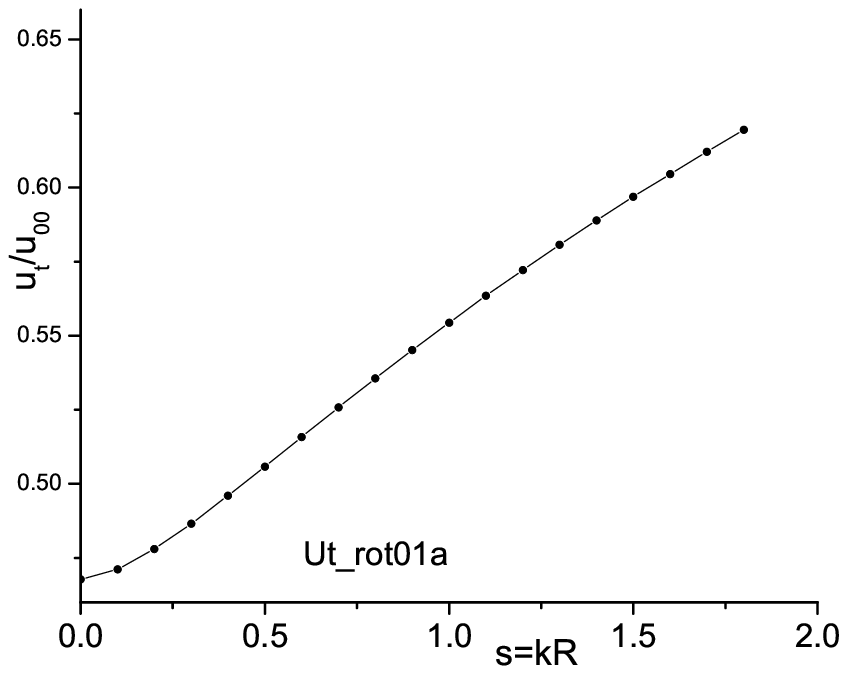}%
}
$

$%
\begin{tabular}
[c]{l}%
Fig.4: The total energy density for a spacial\\
rotating magnetization as function of $s=kR$.
\end{tabular}
\ $%
\[
\]

Finally we want to discuss the physics of the solution and compare it with the experiment.

The numerical results yield a rather short period for the modulation, about 3
times the radius. The reason for this short period is the smallness of the
exchange stiffness. The sum of the demagnetization EDs $\left(  u_{x}%
+u_{z}\right)  /u_{00}$ decreases monotonically with increasing $s=kR$ (for
constant $\theta_{0}$). Only the exchange stiffness which increases as $k^{2}$
can limit the value for $s$.

In the real world the modulation of the magnetization has to overcome a
serious obstacle, the pinning forces in the wire. The coercitive force is a
manifestation of such pinning forces. In future work we intend to determine
the strength of the nucleation force for this modulation. It has to be
stronger than the pinning force to achieve the periodic structure. However,
there are a number of MFM images which show a quasi-period modulation of the
magnetic field along the Co NW. In ref \cite{H39}, Fig.12, two MFM images are
shown of a Co NW which is touched by a short NW. The images appear to show a
periodic sequence of light and dark spot (in the densimeter trace along the NW
does not resolve the fine structure). In ref. \cite{G54} the MFM image of a Co
NW with $2R=35nm$ shows a quasi-periodic field. However, the ratio of period
to radius is not easily extracted from these images. One particularly good
example are the experiments by Belliard et al. \cite{B171} with [Co/Cu] NWs.
For example MFM images of a multi-wire with [$170nm$ Co/$10nm$ Cu] appear to
show opposite magnetization for neighboring segments. We expect that the
demagnetization ED causes an anti-ferromagnetic coupling between neighboring
Co segments.

It throws some additional light on the physics of the modulated magnetization
if one applies the above considerations to a Co wire with a macroscopic
radius, for example $2R=.8mm$. If one assumes as before a modulation of
$\theta=\theta_{0}\cos\left(  kz\right)  $ then one obtains an optimal ED of
$u_{t}/u_{00}=0.213\,62$. This is about half the energy for the magnetization
parallel to the wire axis. However, for a macroscopic wire one should replace
the sinusoidal phase modulation by a more favorable one, close to a square
wave. This will reduce the ED even further. This calculation is in progress
and will be published elsewhere.

It is quite remarkable that we learn from the study of nanowires that the
classical "ground state" of a macroscopic uniaxial wire is very different from
what we thought it was. Of course, in the real world it will be very hard to
prepare a macroscopic Co wire with sufficiently small concentration of pinning
centers so that the magnetization can optimally align. Nanowires are much
better suited for the observation of this modulation because they have fewer
pinning centers.

Abbreviations used: NW = nanowire, ED = energy density.%

\[
\]

\end{document}